\documentclass[a4paper]{spie}  

\usepackage{amsmath,amsfonts,amssymb}
\usepackage{textcomp}
\usepackage{graphicx,subfig}
\usepackage[colorlinks=true, allcolors=blue]{hyperref}

\title{A universal smartphone add-on for portable spectroscopy and polarimetry: iSPEX 2}

\author[a,b]{Olivier Burggraaff}
\author[c]{Armand B. Perduijn}
\author[c]{Robert F. van Hek}
\author[d]{Norbert Schmidt}
\author[a]{Christoph U. Keller}
\author[a]{Frans Snik}
\affil[a]{Leiden Observatory, Leiden University, PO Box 9513, 2300 RA Leiden, The Netherlands}
\affil[b]{Institute of Environmental Sciences (CML), Leiden University, PO Box 9518, 2300 RA Leiden, The Netherlands}
\affil[c]{Bright LED Solutions bv, De Run 4348, 5503 LN Veldhoven, The Netherlands}
\affil[d]{DDQ Apps, Webservices, Project Management, Maastricht, The Netherlands}

\authorinfo{Further author information: (Send correspondence to O.B.)\\O.B.: E-mail: burggraaff@strw.leidenuniv.nl}

\begin{document} 
\maketitle

\begin{abstract}
Spectropolarimetry is a powerful technique for remote sensing of the environment. It enables the retrieval of particle shape and size distributions in air and water to an extent that traditional spectroscopy cannot. SPEX is an instrument concept for spectropolarimetry through spectral modulation, providing snapshot, and hence accurate, hyperspectral intensity and degree and angle of linear polarization. Successful SPEX instruments have included groundSPEX and SPEX airborne, which both measure aerosol optical thickness with high precision, and soon SPEXone, which will fly on PACE. Here, we present a low-cost variant for consumer cameras, iSPEX 2, with universal smartphone support. Smartphones enable citizen science measurements which are significantly more scaleable, in space and time, than professional instruments. Universal smartphone support is achieved through a modular hardware design and SPECTACLE data processing. iSPEX 2 will be manufactured through injection molding and 3D printing. A smartphone app for data acquisition and processing is in active development. Production, calibration, and validation will commence in the summer of 2020. Scientific applications will include citizen science measurements of aerosol optical thickness and surface water reflectance, as well as low-cost laboratory and portable spectroscopy.

\end{abstract}


\section{INTRODUCTION} \label{s:intro}

Spectropolarimetry, the characterization of reflected or emitted light at different wavelengths and polarization states, is a powerful technique for remote sensing of the environment \cite{Dubovik2019polarimetryaerosols, Neukermans2018remotesensingreview, Snik2014polreview}. Most prominently, the Plankton, Aerosol, Cloud, ocean Ecosystem (PACE) satellite due for launch in 2022 or 2023 will fly two spectropolarimetric instruments, namely HARP-2 and SPEXone \cite{Werdell2019PACE}. HARP-2 will observe linear polarization (LP) in four spectral bands (440, 550, 670, 870 nm) at 10--60 angles with a polarimetric accuracy of $<0.005$ in Degree of Linear Polarization (DoLP) \cite{Werdell2019PACE, Remer2019PACEaerosolSPEX}. Meanwhile, SPEXone will observe at five discrete angles ($0^\circ$, $\pm 20^\circ$, $\pm 57^\circ$) with continuous spectral coverage from 385--770 nm and a DoLP accuracy of 0.0025 \cite{Hasekamp2019SPEXone, Remer2019PACEaerosolSPEX}. Instruments observing circular polarization are also under active development, such as the Life Signature Detection polarimeter (LSDpol) \cite{Snik2019LSDpol}, but current efforts typically focus on linear polarimetry, as does this work.

Science cases for linear spectropolarimetry include the retrieval of aerosol and hydrosol particle properties, the beam attenuation and absorption coefficients ($c$, $a$) in water, and the study of vegetation covers. For aerosols, there is already a large history of multi-angle spectropolarimetric observations, from which parameters including particle size and shape distributions, spatial distributions, and chemical composition can be derived \cite{Dubovik2019polarimetryaerosols, Remer2019PACEaerosolSPEX}. More recently, this has been extended to oceanic hydrosols, where the bulk refractive index, particle size distribution, and $c$ can be derived from DoLP \cite{Ibrahim2016hydrosolpolarimetry}. This has been demonstrated for example by Gilerson et al. with a retrieval algorithm for $c$ and $a$ from multi-angular DoLP data \cite{Gilerson2020spectropolarimetrywater}. Finally, spectropolarimetry of vegetation probes its physical characteristics, such as leaf orientation, and provides reflectance distribution functions, which are crucial for improving the accuracy of air- or space-based aerosol retrieval algorithms. \cite{Sun2017nenulgs_vegetation}

Combining spectral and polarimetric measurements can be done in multiple ways \cite{Tyo2006polarimetryreview}. First, regular spectro\-radiometers can be fitted with rotating polarizing filters, as was done in the aforementioned studies of water and vegetation \cite{Gilerson2020spectropolarimetrywater, Sun2017nenulgs_vegetation}. A second method is `channeled' spectro\-polarimetry, where polarization information is encoded into the spectrum itself. One method for channeled linear spectropolarimetry is SPEX \cite{Snik2009SPEXconcept}, the basis for SPEXone \cite{Hasekamp2019SPEXone}. In SPEX, incoming light is modulated with a sine wave with an amplitude and phase depending on the DoLP and the Angle of Linear Polarization (AoLP), respectively \cite{Snik2009SPEXconcept}. This is further explained in Sec.~\ref{ss:optics:spex}.

The SPEX technique has been applied successfully in two high-end field-going instruments measuring aerosol optical thickness (AOT, sometimes termed aerosol optical depth, AOD), namely groundSPEX \cite{vanHarten2014groundSPEX} and SPEX airborne \cite{Smit2019SPEXairborne}. GroundSPEX is a ground-based instrument based on a dual-channel fiber-optic spectrometer with SPEX optics on a moving mount, allowing sequential measurements at multiple angles. Its AOT measurements are well-correlated (Pearson $r = 0.932$) \cite{vanHarten2014groundSPEX} with data from AERONET, the global network of photometers observing the solar almucantar and principal plane \cite{Holben1998AERONET}. SPEX airborne, as the name implies, is an airborne instrument, simultaneously observing at nine fixed viewing angles. A 2017 campaign on a NASA ER-2 high-altitude aircraft demonstrated excellent agreement (RMS DoLP differences of 0.004--0.02) with coflying instruments \cite{Smit2019SPEXairborne}.

A third successful SPEX variant was iSPEX, a smartphone-based version \cite{Land2016iSPEX, Snik2014ispex}. Developed as a low-cost citizen science (CS) tool for AOT measurements, in 2013 iSPEX was used in CS campaigns yielding $\sim$10\,000 observations in the Netherlands. iSPEX data agreed well with AERONET reference data, showing typical standard errors and offsets in AOT of $<0.1$, while the typical absolute DoLP uncertainties were $\approx 0.03$ \cite{Snik2014ispex}. However, the original iSPEX add-on, app, and data had several limitations. First and foremost, the add-on was tailored to the iPhone 4 and 5 and did not work on later iPhone models (bar the iPhone SE) or any Android devices, limiting its reach and future compatibility. Second, at the time iOS only offered very limited camera controls for third-party applications, meaning iSPEX spectra were gathered at very coarse resolution, in the highly non-linear JPEG format, and with varying and uncontrollable exposure settings \cite{Burggraaff2019SPECTACLE}. Thus, iSPEX data were only reliable when averaged over at least 50 individual measurements. Finally, iSPEX had a single-beam SPEX implementation, meaning the polarization modulation could not be distinguished from inherent spectral features \cite{Snik2014ispex}.

We present iSPEX 2, an upgraded version of iSPEX, solving the problems faced by its predecessor. First, the iSPEX 2 hardware is designed to universally support all smartphones. Second, using our SPECTACLE method and database for camera calibration, smartphone cameras offer data similar in quality to professional radio\-meters \cite{Burggraaff2019SPECTACLE}. Third, a dual-beam SPEX implementation facilitates distinguishing between spectral and polarimetric signals. This improves the polarimetric accuracy and enables pure spectroscopy for uniform targets.

iSPEX 2 was developed specifically for aerosol and ocean color measurements. Like its predecessor, it can be used for large CS campaigns to measure AOT \cite{Snik2014ispex}, with higher quality data, but also for individual AOT measurements like other SPEX variants \cite{Smit2019SPEXairborne, vanHarten2014groundSPEX}, though with a coarser spectral resolution. These can be used to fill in temporal and spatial gaps in AOT coverage for satellite atmospheric correction algorithms. The AOT data may be further improved through aureole, almucantar, and near-horizon measurements \cite{Vlemmix2010MAX-DOAS, Holben1998AERONET, Deepak1982aureole}. Ocean color measurements will include unpolarized remote sensing reflectance ($R_{rs}$), similar to the HydroColor app \cite{Leeuw2018hydrocolor} but hyperspectral, and polarized reflectance as discussed above \cite{Gilerson2020spectropolarimetrywater, Neukermans2018remotesensingreview}. These too can be used to provide coverage without coverage by high-end sensors, but also to validate satellite measurements. Finally, iSPEX 2 can be used as a low-cost instrument for portable or laboratory spectroscopy \cite{McGonigle2018spectrometers, Crocombe2018spectroscopy}.

The working principle of SPEX and its implementation in iSPEX 2 are described in Sec.~\ref{s:optics}. The physical design of the add-on is described in Sec.~\ref{s:design}. Sec.~\ref{s:production} describes the intended production process. The current and planned data acquisition and processing pipeline are given in Sec.~\ref{s:data}. Finally, Sec.~\ref{s:future} contains future plans for calibration, validation, and scientific applications of iSPEX 2.

\section{WORKING PRINCIPLE} \label{s:optics}

\subsection{Definitions}

Spectral polarization states are most easily described using a wavelength-dependent Stokes vector $\vec S (\lambda)$, defined in Eq.~\eqref{e:stokes}. Here $I (\lambda)$ is the total spectral radiance, $Q (\lambda)$ and $U (\lambda)$ the linear polarization state, and $V (\lambda)$ the circular polarization state. Here, we define $+Q$ as horizontal and $-Q$ as vertical polarization, $+U$ and $-U$ as $+45^\circ$ and $-45^\circ$ from the horizontal, and $+V$ and $-V$ as right- and left-handed circular polarization, respectively. Lowercase $q, u, v$ are the fractional polarization, normalized by $I (\lambda)$. $I, Q, U, V$ are sometimes referred to as $S_0, S_1, S_2, S_3$ respectively \cite{Snik2014polreview}. In this work, circular polarization in incoming light will be neglected as typically $v \lesssim 10^{-3}$ in nature \cite{Snik2019LSDpol}.

\begin{equation} \label{e:stokes}
    \vec S (\lambda) 
    = \begin{pmatrix} I (\lambda) \\ Q (\lambda) \\ U (\lambda) \\ V (\lambda) \end{pmatrix}
    = I (\lambda) \begin{pmatrix} 1 \\ q (\lambda) \\ u (\lambda) \\ v (\lambda) \end{pmatrix}
    = \begin{pmatrix} I_0 (\lambda) + I_{90} (\lambda) \\ I_0 (\lambda) - I_{90} (\lambda) \\ I_{45} (\lambda) - I_{-45} (\lambda) \\ I_R (\lambda) - I_L (\lambda) \end{pmatrix}
\end{equation}

The state of linear polarization is also described by the degree and angle of linear polarization, DoLP or $P_L (\lambda)$ and AoLP or $\phi_L (\lambda)$ respectively. These are defined in Eqs.~\eqref{e:dolp} and \eqref{e:aolp} \cite{Snik2009SPEXconcept}. In practice, the arctan2 operator is used in Eq.~\eqref{e:aolp}.

\begin{equation} \label{e:dolp}
    P_L (\lambda) = \frac{\sqrt{Q(\lambda)^2 + U(\lambda)^2}}{I(\lambda)} = \sqrt{q(\lambda)^2 + u(\lambda)^2}
\end{equation}

\begin{equation} \label{e:aolp}
    \phi_L (\lambda) = \frac{1}{2} \arctan \left( \frac{U(\lambda)}{Q(\lambda)} \right) = \frac{1}{2} \arctan \left( \frac{u(\lambda)}{q(\lambda)} \right)
\end{equation}

Finally, optical elements are described through their $4 \times 4$ Mueller matrix $M$, describing how the element modifies the incident $\vec S (\lambda)$. Each element of $M$ can have its own wavelength dependence. Passing through an element $X$ modifies $\vec S (\lambda)$ to be $M_X \vec S (\lambda)$. The Mueller matrix of a chain of elements $X, Y, Z$ is simply the product of their individual Mueller matrices $M_Z M_Y M_X$.

\subsection{SPEX Polarization Modulation Optics} \label{ss:optics:spex}

The SPEX polarization modulation optics (PMO) consist of three elements, namely a quarter-wave plate (QWP), multi-order retarder (MOR), and analyzing linear polarizer (ALP) \cite{Snik2009SPEXconcept}. Their orientations and function are as follows:

\begin{itemize}
    \item \emph{Quarter-wave plate}: The QWP has its fast axis at $0^\circ$ ($+Q$, horizontal). It should be highly achromatic and interchanges the Stokes $U$ and $V$ components, making the instrument insensitive to circular polarization. Residual chromaticity from misalignment, deviations in retardance $\delta_{QWP} (\lambda)$, and other effects must be calibrated \cite{vanHarten2014groundSPEX}. In iSPEX 2, an Edmund Optics WP140HE (\#88-253) $\lambda/4$ polymer retarder foil is used.
    
    \item \emph{Multi-order retarder}: The MOR has its fast axis at $+45^\circ$ from horizontal ($+U$). Its retardance $\delta_{MOR} (\lambda)$ is highly chromatic, exchanging the incoming $Q$ and $V$ components by a fraction depending on the wavelength. As with the QWP, the performance of the MOR requires extensive calibration \cite{vanHarten2014groundSPEX}. The first iSPEX 2 units contain a stack of two Meadowlark B4 polymer retarder foils \cite{Baur2019foils}, with a nominal retardance of $4 \lambda$ each at 560 nm. For the future, alternatives are being investigated, as described in Sec.~\ref{ss:production:mor}. 
    
    \item \emph{Analyzing linear polarizer}: The ALP imprints the modulation onto the exiting spectrum in Stokes $I$, and can be implemented in several ways. The single-beam approach uses a single linear polarizer, parallel or orthogonal to the slit, as in the original iSPEX \cite{Snik2014ispex}. This approach does not allow for full linear spectro\-polarimetry, as the modulation and inherent spectral properties cannot be fully distinguished. This is possible in the dual-beam approach, where both directions are measured. A polarizing beamsplitter is used in groundSPEX \cite{vanHarten2014groundSPEX}, SPEX airborne \cite{Smit2019SPEXairborne}, and SPEXone \cite{Hasekamp2019SPEXone}. In iSPEX 2, a pair of Polarization.com PFSC NA foils is used, oriented parallel or orthogonal to the two slits (Sec.~\ref{ss:design:tube}).
\end{itemize}

Together, the PMO induce a modulation in the outgoing Stokes $I$ radiance $I_\pm (\lambda)$ (where the sign $\pm$ corresponds to the two ALP orientations) that depends only on the incoming radiance $I_{in} (\lambda)$, $P_L (\lambda)$, $\phi_L (\lambda)$, $\lambda$, and $\delta_{MOR} (\lambda)$. This is described in Eq.~\eqref{e:spex}. The modulation is a sine wave on the radiance spectrum, quasi-periodic in $1/\lambda$, its amplitude and phase corresponding to the DoLP ($P_L$) and AoLP ($\phi_L$), respectively \cite{Snik2009SPEXconcept}. These parameters are retrieved by fitting Eq.~\eqref{e:spex}.

\begin{equation} \label{e:spex}
    I_\pm (\lambda) = \frac{I_{in} (\lambda)} {2} \left[ 1 \pm P_L (\lambda) \cos \left( \frac{2 \pi \delta_{MOR} (\lambda)}{\lambda} + 2 \phi_L (\lambda) \right) \right]
\end{equation}

In dual-beam mode, the modulations in $I_+ (\lambda)$ and $I_- (\lambda)$ are exactly opposite for a uniform target, so the total radiance and modulation can be disentangled as shown in Eqs.~\eqref{e:spex_I} and \eqref{e:spex_mod}. However, this is complicated in practice due to imperfections in the optical elements, misalignments, differences in transmission between the two beams, and nonnormal incidence \cite{Smit2019SPEXairborne, vanHarten2014groundSPEX}.

\begin{equation} \label{e:spex_I}
    I_+ (\lambda) + I_- (\lambda) = I_{in} (\lambda)
\end{equation}

\begin{equation} \label{e:spex_mod}
    \frac{I_+ (\lambda) - I_- (\lambda)}{I_+ (\lambda) + I_- (\lambda)} = P_L (\lambda) \cos \left( \frac{2 \pi \delta_{MOR} (\lambda)}{\lambda} + 2 \phi_L (\lambda) \right)
\end{equation}

\section{ADD-ON DESIGN} \label{s:design}

\begin{figure}[ht]
    \centering
    \subfloat[iSPEX 2, whole. \label{f:ispex_exploded:whole}]{\includegraphics[height=6.4cm]{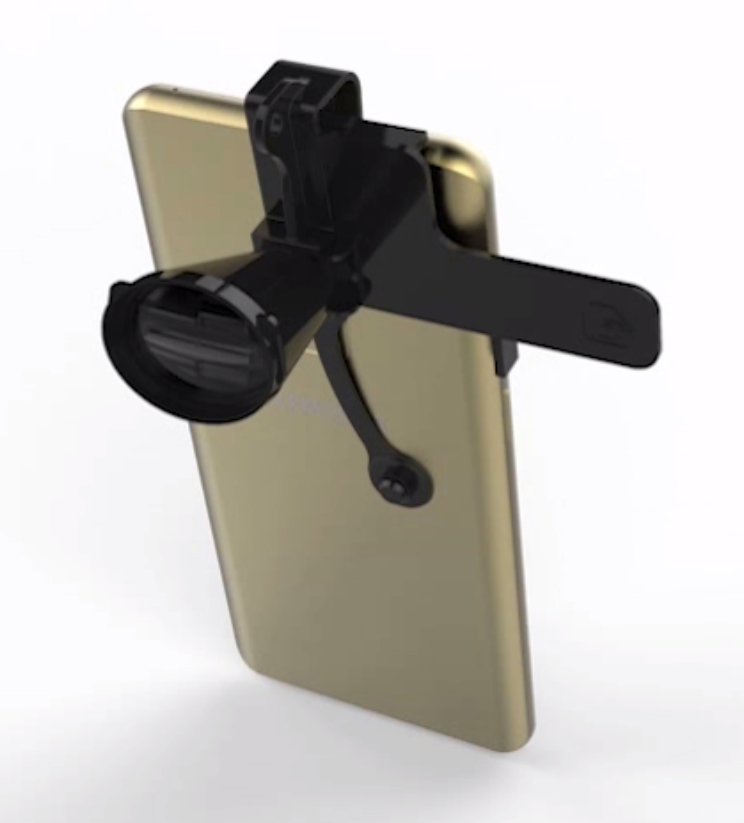}}
    \qquad
    \subfloat[iSPEX 2, exploded view. \label{f:ispex_exploded:exploded}]{\includegraphics[height=6.4cm]{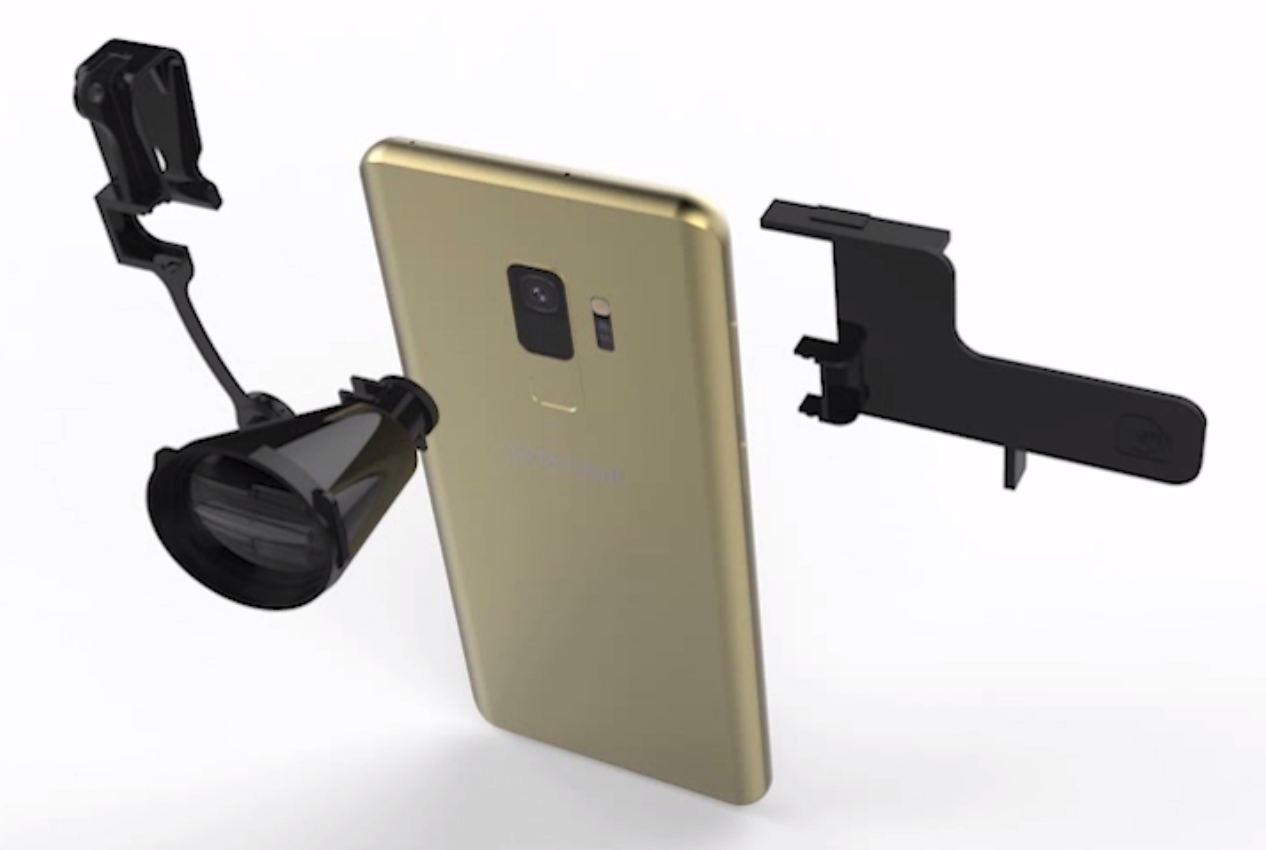}}
    \caption{Render of the iSPEX 2 add-on, as a whole attached to a smartphone (left) and exploded into its three components (right). These are, from left to right, the smartphone clip, optical tube, and smartphone backplate. The smartphone is seen from the back.}
    \label{f:ispex_exploded}
\end{figure}

\begin{figure}[ht]
    \centering
    \includegraphics[width=\textwidth]{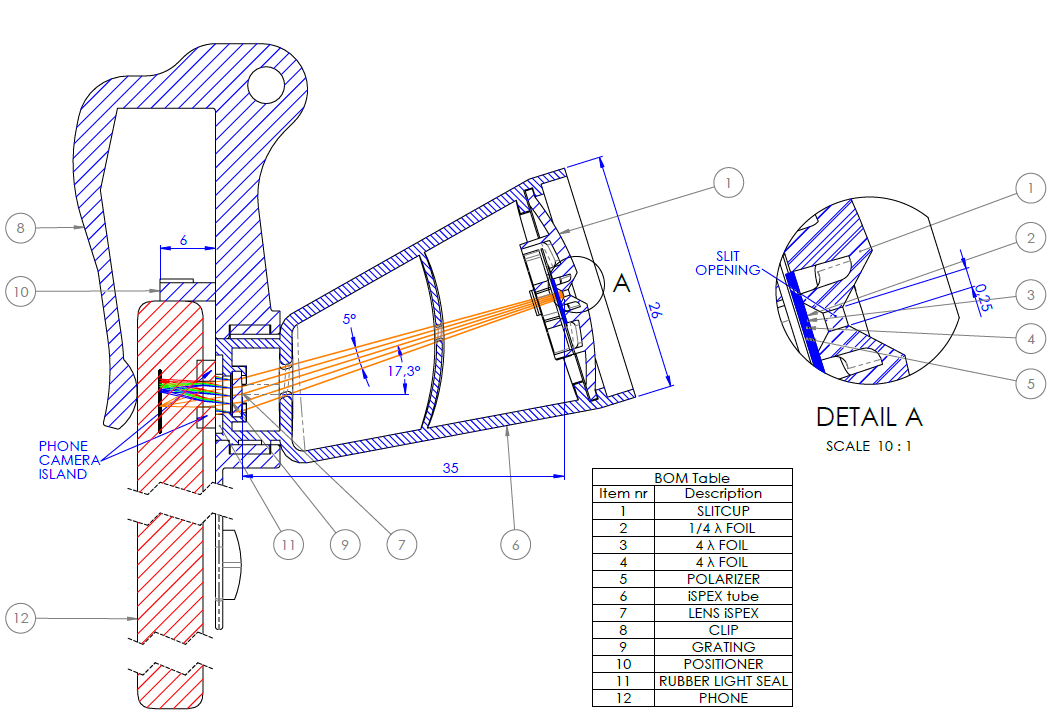}
    \caption{Cross section of iSPEX 2 (blue) attached to a smartphone (red) with a zoom on the slit (detail A). Five rays are traced from the slit to the grating in orange, then dispersed onto the camera chip. Sizes are in mm.}
    \label{f:cad_design}
\end{figure}

The iSPEX 2 add-on is a whole divided into three parts, as shown in Fig.~\ref{f:ispex_exploded}. These are a tube containing the PMO and other optics (Sec.~\ref{ss:design:tube}), a clip to clamp onto a smartphone (Sec.~\ref{ss:design:clip}) and a backplate for aligning the tube with the smartphone camera (Sec.~\ref{ss:design:backplate}). A cross-section is shown in Fig.~\ref{f:cad_design}. The tube can be used on any camera, including smartphones but also UAVs and webcams. The clip can be used with nearly all smartphones, as most models have a similar form factor \cite{Burggraaff2019SPECTACLE}. Finally, the backplate is unique to each smartphone model.

\subsection{Optical Tube} \label{ss:design:tube}

\begin{figure}[ht]
    \centering
    \includegraphics[width=0.6\textwidth]{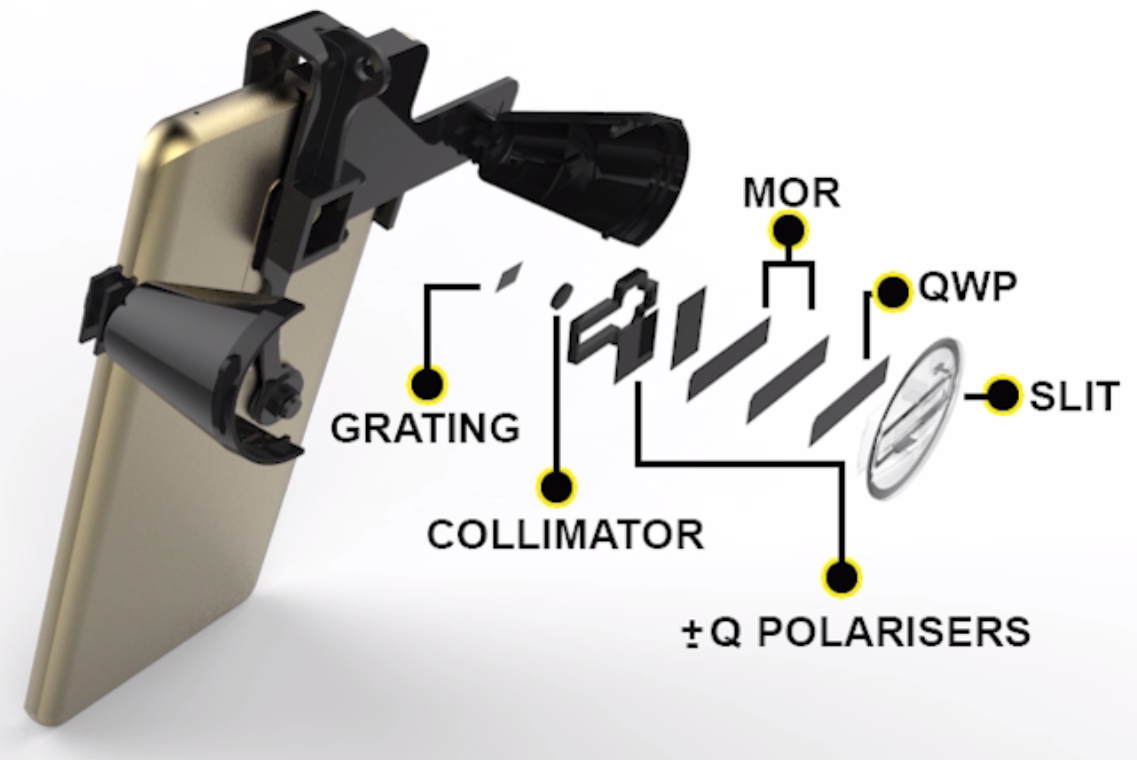}
    \caption{Exploded view of the iSPEX 2 optical tube with the slit, quarter-wave plate (QWP), multi-order retarder (MOR), $0^\circ$ and $90^\circ$ polarizers, collimator lens, and grating foil indicated.}
    \label{f:optics_exploded}
\end{figure}

The iSPEX 2 optics consist of a double slit (side by side), the SPEX PMO (Sec.~\ref{ss:optics:spex}), a collimator lens, and a transmission grating, as shown in Fig.~\ref{f:optics_exploded}. There are two slits, each 0.25 mm wide and 9 mm long, located side by side to measure in quasi-dual-beam mode. The PMO are placed directly behind the slits to minimize instrumental polarization through stray light. A small plastic cradle holds the PMO foils in place. 
Dual-beam mode is achieved by having a horizontal ALP in the PMO behind the left slit and a vertical one behind the right slit. Dual-beam mode requires a uniform target between the two slits; this assumption holds in the center for smooth surfaces like sky polarization \cite{vanHarten2014groundSPEX}, but not toward the edges.
From the PMO, the modulated light propagates to a custom-made collimator lens ($f = 35$ mm) and a 1000 line/mm holographic transmission grating foil (Edmund Optics \#52-116), dispersing the light onto the smartphone camera; the camera optics then register the spectra, as shown in Figs.~\ref{f:cad_design} and \ref{f:raw_data}. A rubber seal blocks stray light, as shown in Fig.~\ref{f:cad_design}.

The optics are located in a plastic tube, as shown in Fig.~\ref{f:optics_exploded}. The tube itself is 35 mm long along its optical axis, which is angled $+17.3^\circ$ upwards to project the entire zeroth and first orders of the spectrum on the smartphone camera. The tube consists of two halves (left and right) which are produced separately and click together along its length (Sec.~\ref{ss:production:assembly}). A baffle is located halfway along the tube, consisting of overlapping protrusions from either tube half. This ensures overlapping coverage and thus reduces light leakage. The slit end of the tube has two `ears' to which additional add-ons can be attached, such as a cuvette holder for transmission spectroscopy. The camera end has two ridges to which the clip (Sec.~\ref{ss:design:clip}) attaches; these can also be used for custom attachments for different cameras.

\subsection{Smartphone Clip} \label{ss:design:clip}

\begin{figure}[ht]
    \centering
    \subfloat[Smartphone clip.]{\includegraphics[height=8.8cm,angle=90]{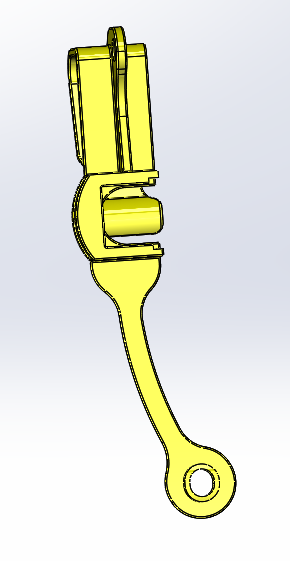}}
    \qquad
    \subfloat[Zoom on clamp.]{\includegraphics[width=5cm]{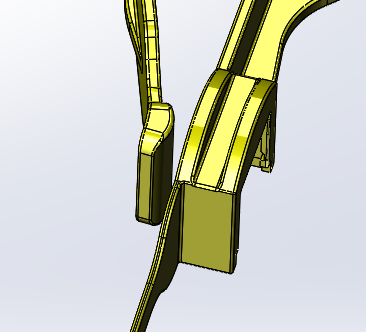}}
    \caption{CAD model of the iSPEX 2 smartphone clip. The optical tube and backplate slot into the central opening, opposite the clamp. The long, curved extension is the `clapper' with a suction cup at the end. Ridges along the clip provide stiffness and strength. The clip is fabricated in black.}
    \label{f:clip}
\end{figure}

iSPEX 2 attaches to smartphones using a clip, as seen in Fig.~\ref{f:ispex_exploded}. The clip design is shown in detail in Fig.~\ref{f:clip}. It attaches to the smartphone with a clamp on the front (screen) side, directly behind the camera. This clamp is 14 mm wide and made from soft plastic to prevent scratching. It is attached to the clip over the top of the smartphone. Additionally, the clip has a `clapper' extending 49 mm below the camera along the back side, with a suction cup at the end. This attaches to the flat back surface of the smartphone. This double attachment prevents rotation of the add-on and ensures the slit is always projected horizontally onto the camera. The clapper is curved 11 mm to the right of the camera (seen from the back, as shown in Fig.~\ref{f:backplate_5phones}) so the suction cup does not fall off the edge on devices with cameras near the edge, such as iPhones.

Several iterations of the clip design were necessary. Originally, the clip had no clapper and attached to the right side of the smartphone (seen from the front), but this acted as a lever and caused the add-on to rotate under its own weight. This was solved by having the clip attach over the top, so its weight rests on the smartphone, and adding a clapper. The clapper originally had a small magnet at its end instead of a suction cup, but smartphone backsides were found to be only weakly magnetic in only a few places.

\subsection{Smartphone Backplate} \label{ss:design:backplate}

\begin{figure}[ht]
    \centering
    \includegraphics[width=\textwidth]{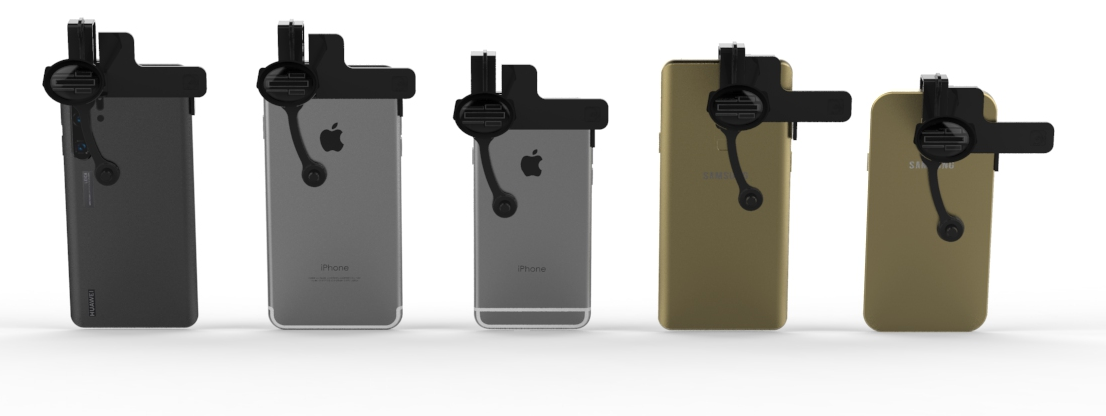}
    \caption{Render of iSPEX 2 attached to several smartphones with different dimensions and camera locations, seen from the back. Each has a backplate with unique positioner locations to place the optics directly in front of the camera.}
    \label{f:backplate_5phones}
\end{figure}

iSPEX 2 is placed in front of the smartphone camera using a backplate unique to each smartphone model, as shown in Fig.~\ref{f:backplate_5phones}. Smartphones with multiple backside cameras typically feature one wide-view camera with generic optical properties and a focal length of 3.8--4.5 mm (corresponding to a field-of-view of 60--75$^\circ$ $\times$ 45--55$^\circ$ \cite{Burggraaff2019SPECTACLE}, which is used for iSPEX 2. The backplate consists of a universal plate with two positioners with different positions for each smartphone model. One rests on top of the smartphone, the other on the right (seen from the back). Since most smartphone cameras are in the center or on the left, this design ensures that most of the weight of iSPEX 2 rests on the smartphone, rather than create a lever. 

The positioner locations depend on the dimensions of the smartphone, the curvature of its top corners, and the locations of buttons along the side. We are compiling a database of popular smartphone models including these parameters. The top positioner has a small notch to accommodate the iPhone SE, one of the few popular models with buttons along the top. The positioners are 21.5 mm (side) and 30 mm (top) long and 6 mm wide, providing sufficient coverage even on smartphones with curved edges.

\section{PRODUCTION} \label{s:production}

\subsection{Multi-Order Retarder Foils} \label{ss:production:mor}

As described in Sec.~\ref{ss:optics:spex}, iSPEX 2 uses polymer retarder foils to produce the SPEX polarization modulation. The first units use two Meadowlark B4 foils \cite{Baur2019foils} with $4 \lambda$ retardance (at 560 nm) each, as did the original iSPEX \cite{Snik2014ispex}, for a nominal total retardance of $8 \lambda$ at 560 nm. This induces a modulation with 7 full periods across the typical spectral range of smartphone cameras (as seen in Fig.~\ref{f:wavelength:100}), which is 390--700 nm \cite{Burggraaff2019SPECTACLE}.

These foils are produced by stretching transparent sheets of polymer, such as polyvinyl alcohol (PVA), polycarbonate (PC) or poly(methyl methacrylate) (PMMA) \cite{Baur2019foils}. While the pre-fabricated foils provide the desired retardance with great consistency, they are prohibitively expensive for low-cost CS purposes.

To enable high-volume throughput, integrated in our production line, we are exploring internal production of MOR foils, based on a setup previously used to stretch sheet metal \cite{Perduijn1995bendingsheet}. Initial experiments are focused on finding the optimal material from PVA, PC, PMMA, polyethylene terephthalate (PET), and Zeonor cyclo olefin polymer, at various thicknesses from 50--200 \textmu m. Further experiments will determine the efficacy of softening the foils through heat (up to 80~$^\circ$C) and the maximum achievable retardance. Visual inspection and spectrally resolved measurements through a crossed polarizer setup will be used to measure the retardance of sections of the foil during the stretching process, as spatial variations in retardance and fast axis orientation are expected \cite{Baur2019foils}. The end goal is to mass produce low-cost MOR foils with sufficient quality for iSPEX 2, not necessarily for high-end commercial purposes.

\subsection{Injection Molding} \label{ss:production:injection}

Like the original \cite{Snik2014ispex}, iSPEX 2 will be produced through injection molding. This is inexpensive yet precise. Components that can be injection molded include the collimator lens, optical tube, smartphone clip, and backplate. The suction cup (Sec.~\ref{ss:design:clip}) and optical foils (Secs.~\ref{ss:optics:spex} and \ref{ss:design:tube}) are purchased or self-produced by other means (Sec.~\ref{ss:production:mor}).

Various plastics will be used for iSPEX 2. Thin parts, including the tube and backplates, will be manufactured from polycarbonate-acrylonitrile butadiene styrene (PC-ABS) colored black with masterbatch (MB). PC-ABS and polypropylene (PP) are being evaluated as materials for the clip. The collimators are produced separately from Zeonor 330 plastic.

Molds for the optical tube and clip are being manufactured, while a mold for the backplate is still being designed. The backplate mold is complex, requiring two sliders to account for the device-dependent positioners. By parameterizing the slider positions based on the smartphone dimensions, the mold can instantly be adjusted to a different device.

\subsection{3D Printing} \label{ss:production:print}

Except for the optical components, iSPEX 2 units can also be 3D printed, for which we will provide model files. This was used extensively in the development phase for quick testing and will be useful for future compatibility. For example, this allows users to self-produce backplates tailored for new smartphone models not included in our database. Local production through 3D printing also reduces the unit cost, especially valuable in resource-poor areas, one of the prime target audiences for smartphone spectroscopy \cite{McGonigle2018spectrometers}.

However, 3D printing introduces several difficulties. First of all, the PMO and grating foils cannot be 3D printed and thus must still be purchased and cut to size. Second, low-cost 3D printing techniques inherently have wider production tolerances than injection molding. With 3D printed prototypes we often found it necessary to manually file or cut components to make them fit tightly. Finally, some 3D printing materials such as PA nylon are translucent; we found this easiest to counteract by covering the entire unit twice over with a felt-tip pen. Because of these complications, we use 3D printing only for prototyping, not production.

\subsection{Assembly} \label{ss:production:assembly}

Assembly of iSPEX 2 units is straight-forward. First, the PMO foils are placed in their cradle in the correct orientations. Next, the cradle, collimator lens, and grating foil are placed into the corresponding slots in the tube halves, which are then clicked together. When used with a smartphone, the optical tube is then slotted into the smartphone clip, followed by the backplate corresponding to the smartphone model. The backplate is easily removed for use with a different device. For other uses, a custom attachment between tube and camera can easily be manufactured by the user. The tube can even be used with the naked eye for a qualitative measurement or a demonstration.

For the injection molded units, the PMO and grating foils are punched into an asymmetric shape to prevent confusion of their optical axes; the cradle has corresponding protuberances. Like with the original iSPEX, a calibration setup in the factory is used to verify that all optics are oriented correctly \cite{Snik2014ispex}. For 3D printed units, this must be done carefully by hand.

\section{DATA ACQUISITION AND PROCESSING} \label{s:data}

This section describes the acquisition, calibration, and processing of iSPEX 2 data. These are acquired as RAW images on a smartphone using an app based on SPECTACLE \cite{Burggraaff2019SPECTACLE}, described in Sec.~\ref{ss:data:app}. Currently, these RAW data are manually uploaded to a PC for calibration and processing. Our goal is to move as much of this as possible to the smartphone, possibly with additional cloud computing in a back-end server for devices with insufficient computational power.

\subsection{Smartphone App} \label{ss:data:app}

\begin{figure}[ht]
    \centering
    \subfloat[Fluorescent light. \label{f:raw_data:TL}]{\includegraphics[width=8.2cm,height=6.16cm]{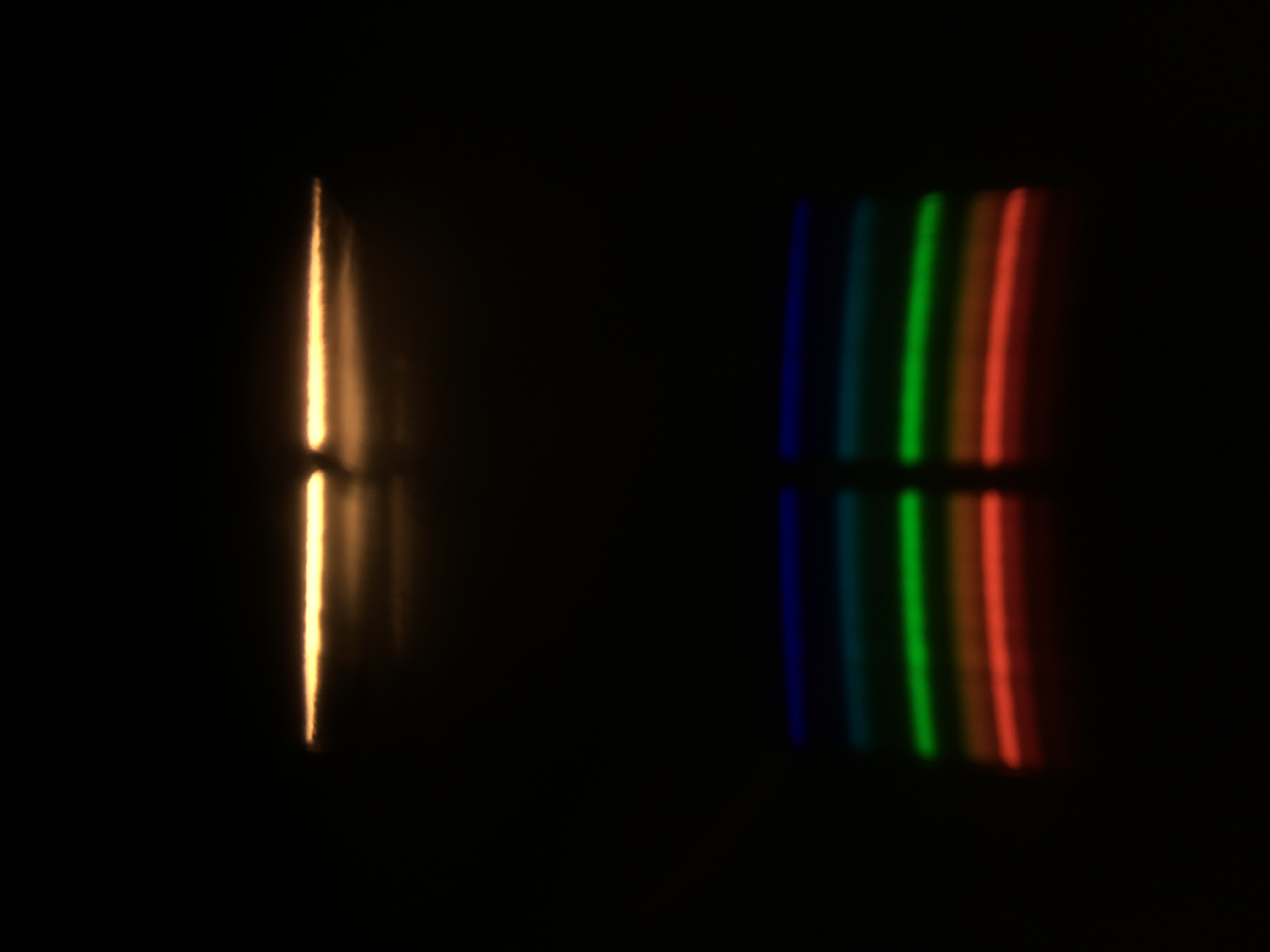}}
    \qquad
    \subfloat[Reflected sunlight through a 100\% polarizer. \label{f:raw_data:100}]{\includegraphics[width=8.2cm,height=6.16cm]{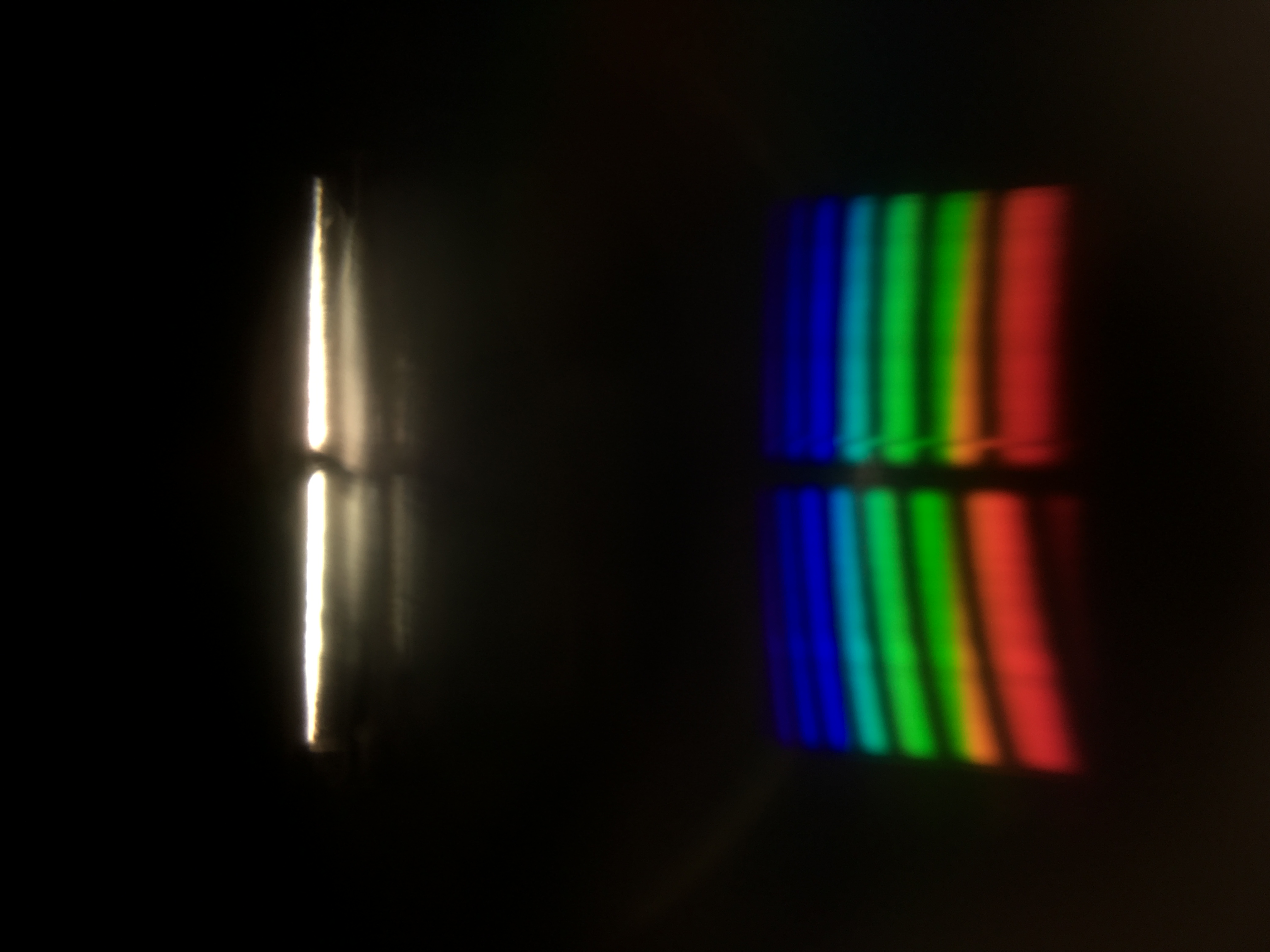}}
    \caption{Spectra taken with an iSPEX 2 prototype on an iPhone SE. The fluorescent light spectrum (left) is used in the wavelength calibration, and clearly shows the smile and keystone effects. The 100\% polarized image (right) shows the SPEX modulation in the two $\pm Q$ spectra, with an additional curvature due to variations in retardance. The small glitch in the top spectrum and the stray light are due to manufacturing faults in this prototype. These images were taken in RAW format and converted to JPEG for visualization; all data processing is done on RAW images only.}
    \label{f:raw_data}
\end{figure}

Data acquisition on smartphones is done using a custom-designed app, based on SPECTACLE \cite{Burggraaff2019SPECTACLE}, currently in development for iOS and Android. Significant changes to iOS mean few elements from the 2013 iSPEX app \cite{Snik2014ispex} can be reused for iSPEX 2. However, user feedback on the original app is taken into account. For example, some users misunderstood the scientific aims and methods of iSPEX because these were not explained clearly \cite{Land2016iSPEX}. Difficulties in installing the add-on and interpreting feedback from the app were also noted \cite{Budde2017participatorysensing}.

Data are obtained in RAW image format because of its high linearity and dynamic range \cite{Burggraaff2019SPECTACLE}. This is in contrast to the JPEG images taken with the original, where nonlinearity and white balance introduced significant problems \cite{Burggraaff2019SPECTACLE, Snik2014ispex}. Two examples, taken with a 3D printed prototype, are shown in Fig.~\ref{f:raw_data}. Aside from problems due to faults in the prototype, such as stray light, which will be reduced in the final product, these images are representative examples of iSPEX 2 data. Processing of these images (Sec.~\ref{ss:data:processing}) is currently done on PC, but will be done in-app in the future.

The following data acquisition protocols will be included in the initial release of the app:

\begin{itemize}
    \item Wavelength calibration: Single observation of a fluorescent light, as described in Sec.~\ref{ss:data:wavelength}.
    \item Aerosols (AOT): Series of observations from horizon to zenith along the principal (observer-Sun-zenith) plane, as with other SPEX variants \cite{Smit2019SPEXairborne, Snik2014ispex, vanHarten2014groundSPEX}. The optimal number of observations is to be determined based on data quality and computational considerations, specifically the speed at which RAW images can be saved on smartphones.
    \item Water ($R_{rs})$: Series of observations according to the Mobley protocol \cite{Mobley1999Rrs}, measuring sky radiance at $40^\circ$ from zenith, upwelling radiance at $40^\circ$ from nadir, and downwelling irradiance with a grey card at $40^\circ$ from nadir. The same protocol is used in the HydroColor app, which does multispectral (RGB) measurements \cite{Leeuw2018hydrocolor}.
\end{itemize}

Users are guided through these protocols with text explaining what to do and, for example, arrows to guide them in the right direction, using the smartphone compass and accelerometers. Citizen scientists have been involved in the development of these protocols from the start, to ensure user-friendliness. Optimal exposure settings for each protocol are currently hard-coded but in the future will be determined automatically.

\subsection{Data Pre-Processing} \label{ss:data:processing}

\begin{figure}[ht]
    \centering
    \subfloat[Fluorescent light. \label{f:raw_lines:TL}]{\includegraphics[width=8.2cm]{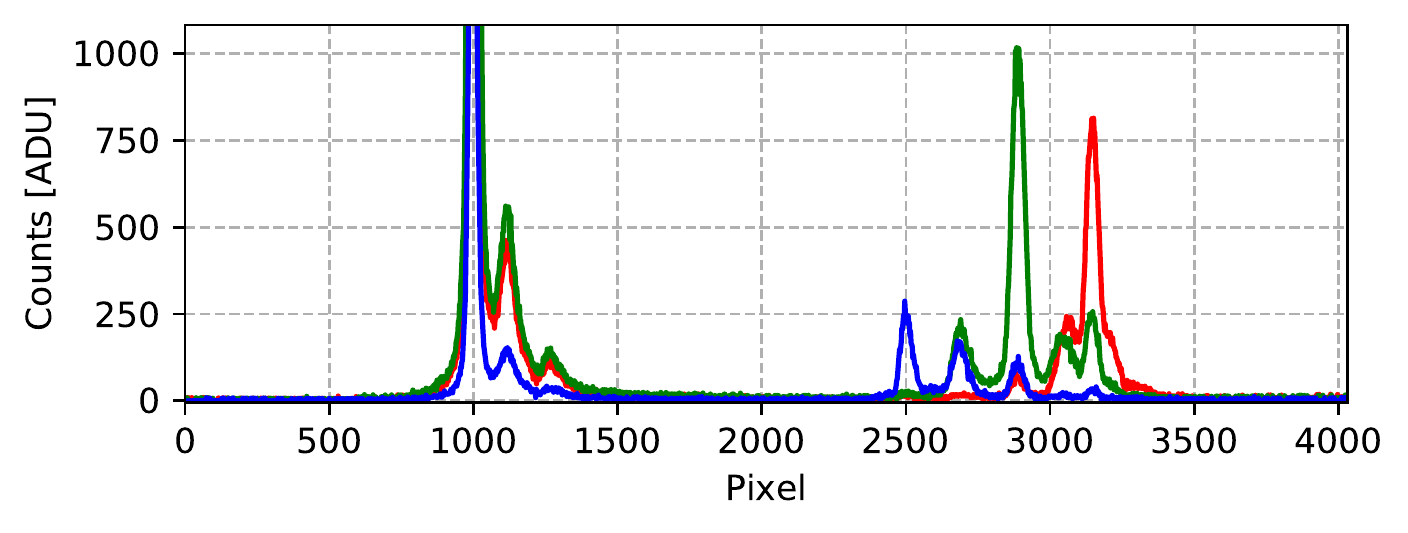}}
    \qquad
    \subfloat[Reflected sunlight through a 100\% polarizer. \label{f:raw_lines:100}]{\includegraphics[width=8.2cm]{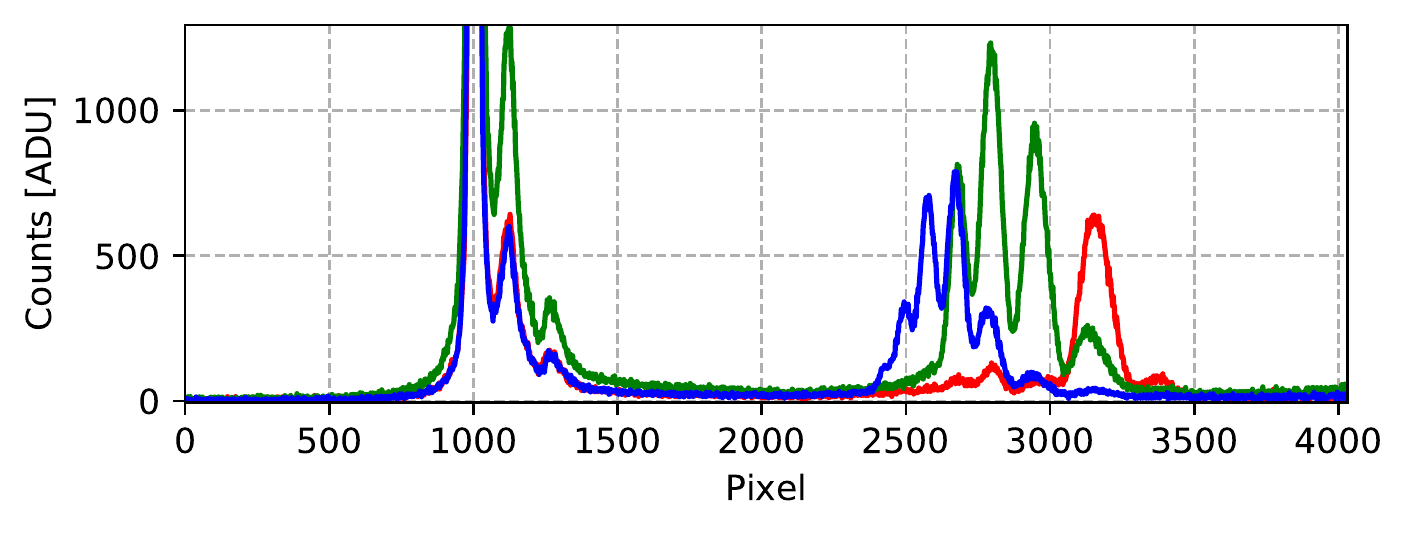}}
    \caption{Two adjacent rows of pixels (one GR, one BG$_2$) in the spectra shown in Fig.~\ref{f:raw_data}, split into the RGB channels. The G$_2$ channel is not shown here. The slit and several ghosts are visible from 800--1500 pixels, the first order spectrum from 2300--3500 pixels. The data have been corrected for bias and flat-field.}
    \label{f:raw_lines}
\end{figure}

Data are processed according to the SPECTACLE method \cite{Burggraaff2019SPECTACLE}, which was originally developed for iSPEX 2. A Python library for data processing specific to iSPEX 2 is currently in development. iSPEX 2 data are corrected for bias and flat-field using SPECTACLE; in spectra taken with prototypes, dark current is negligible compared to stray light, but this may change with injection molded iSPEX 2 units or on certain devices.

The corrected image is split into the $\pm Q$ component spectra, currently based on hard-coded windows for specific cameras but in the future automatically, and demosaiced. The RGBG$_2$ channels of the Bayer-filter camera are treated separately rather than combined through interpolation, since interpolated data add no extra information \cite{Burggraaff2019SPECTACLE}. After demosaicing, there are eight separate spectra, namely the combinations of $\pm Q$ and RGBG$_2$. Fig.~\ref{f:raw_lines} shows two examples of demosaiced RGB spectra. Finally, each row is convolved with a Gaussian kernel ($\sigma = 6$ pixels) corresponding to an FWHM of $\sim$3.8 nm (see Sec.~\ref{ss:data:wavelength}). This reduces the noise on the spectrum without reducing the spectral resolution, being narrower than the image of the slit.

\subsection{Wavelength Calibration} \label{ss:data:wavelength}

The pre-processed data are then wavelength-calibrated. As seen in Fig.~\ref{f:raw_data}, iSPEX 2 data exhibit significant smile (variations in dispersion along the slit) and keystone (deformation of the spectrum into a trapezoid). Both effects are common in long-slit spectrometers \cite{Smit2019SPEXairborne, Yuan2019grismspectrometer}. Smile is corrected by doing the wavelength calibration per pixel row; a keystone correction is still in development.

\begin{figure}[ht]
    \centering
    \includegraphics[width=\textwidth]{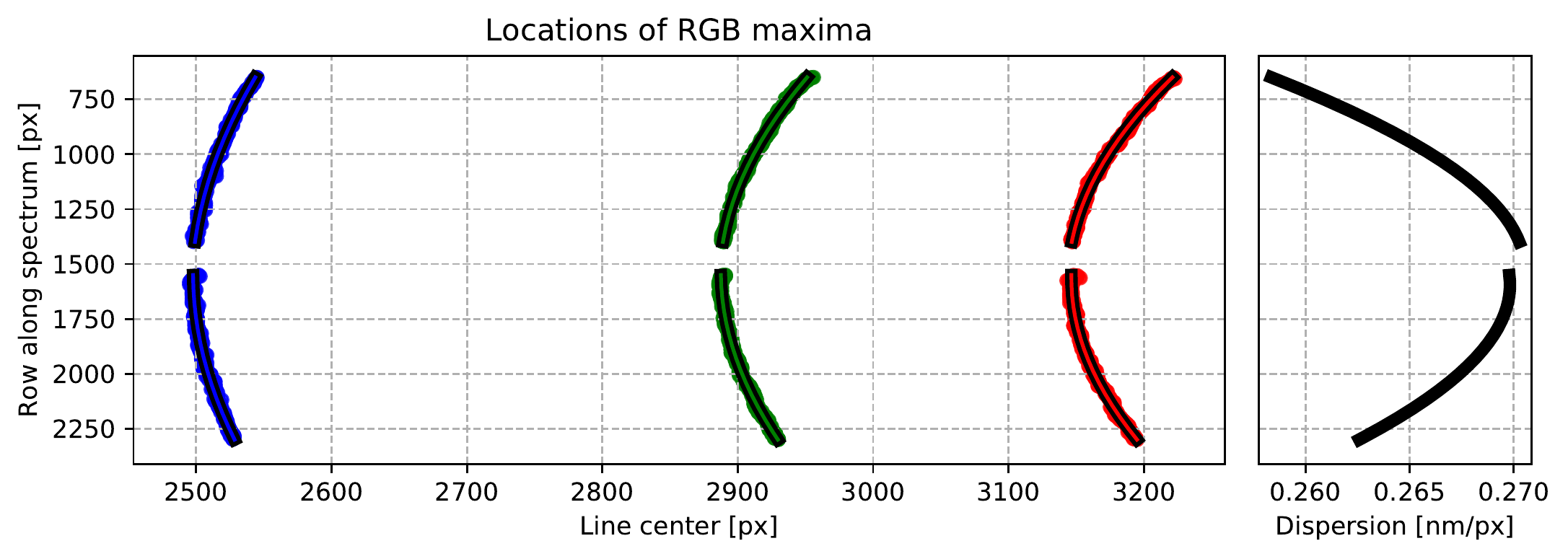}
    \caption{Pixel positions of the spectral lines in Fig.~\ref{f:raw_data:TL} (left) and the derived dispersion (right). The colored dots indicate the maxima in the B, G/G$_2$, R (from left to right) channels, while the colored lines with a black outline indicate the fitted positions. The fits are done separately for the $\pm Q$ spectra. The smile effect is clearly visible. The mean dispersion between the B (436.6 nm) and R (611.6 nm) lines is shown; a smaller value in nm/px corresponds to a wider dispersion.}
    \label{f:wavelength_calibration}
\end{figure}

The wavelength calibration is done using a reference spectrum of a fluorescent light, like that shown in Fig.~\ref{f:raw_data:TL}. These have three sharp spectral lines corresponding to the RGBG$_2$ channels, at 611.6 (R), 544.5 (G/G$_2$), and 436.6 (B) nm. Because of mosaicing, the raw data have R and B values only in every alternate row and column, while combining G and G$_2$ gives full row coverage but in alternating columns. The column gaps are filled in by the Gaussian kernel convolution (Sec.~\ref{ss:data:processing}), which also reduces noise. The maximum value per channel in each pixel row is determined to find the line centers. A quadratic fit is made to these, and the resulting fitted line centers are used. This fills in the gaps in R and B and reduces the effect of noisy rows. Fig.~\ref{f:wavelength_calibration} shows the line centers thus derived from the spectrum in Fig.~\ref{f:raw_data:TL}.

A wavelength solution map, with the central wavelength for each pixel, is generated by fitting a quadratic relation between the spectral line wavelengths and the line centers from Fig.~\ref{f:wavelength_calibration}.  The resulting dispersion is typically $\sim$0.27 nm/px, depending on the camera optics, exposure settings, and pixel position (as seen in Fig.~\ref{f:wavelength_calibration}). For the images shown in Fig.~\ref{f:raw_data}, where the slit is $\sim$35 pixels wide (FWHM), this gives a spectral resolution (FWHM) of 9 nm. The FWHM varies slightly based on camera optics and exposure settings, mainly focus. 9 nm resolution is comparable to common ocean color sensors such as the TriOS RAMSES and HyperOCR\cite{Burggraaff2020convolution} and satellite instruments like Sentinel-3/OLCI \cite{Burggraaff2020convolution} and HARP-2 \cite{Werdell2019PACE}, and only 2--3 times wider than PACE/OCI (5 nm) \cite{Werdell2019PACE} and SPEXone (2--3 nm) \cite{Werdell2019PACE, Hasekamp2019SPEXone}.

\begin{figure}[ht]
    \centering
    \subfloat[Fluorescent light. \label{f:wavelength:TL}]{\includegraphics[width=8.2cm]{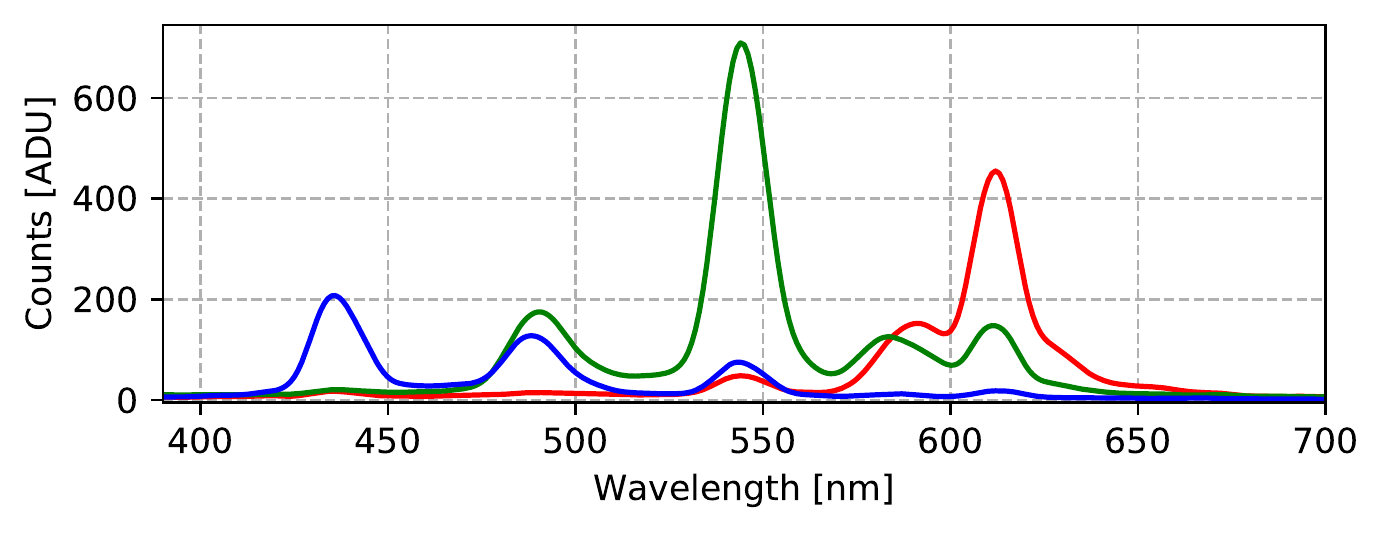}}
    \qquad
    \subfloat[Reflected sunlight through a 100\% polarizer. \label{f:wavelength:100}]{\includegraphics[width=8.2cm]{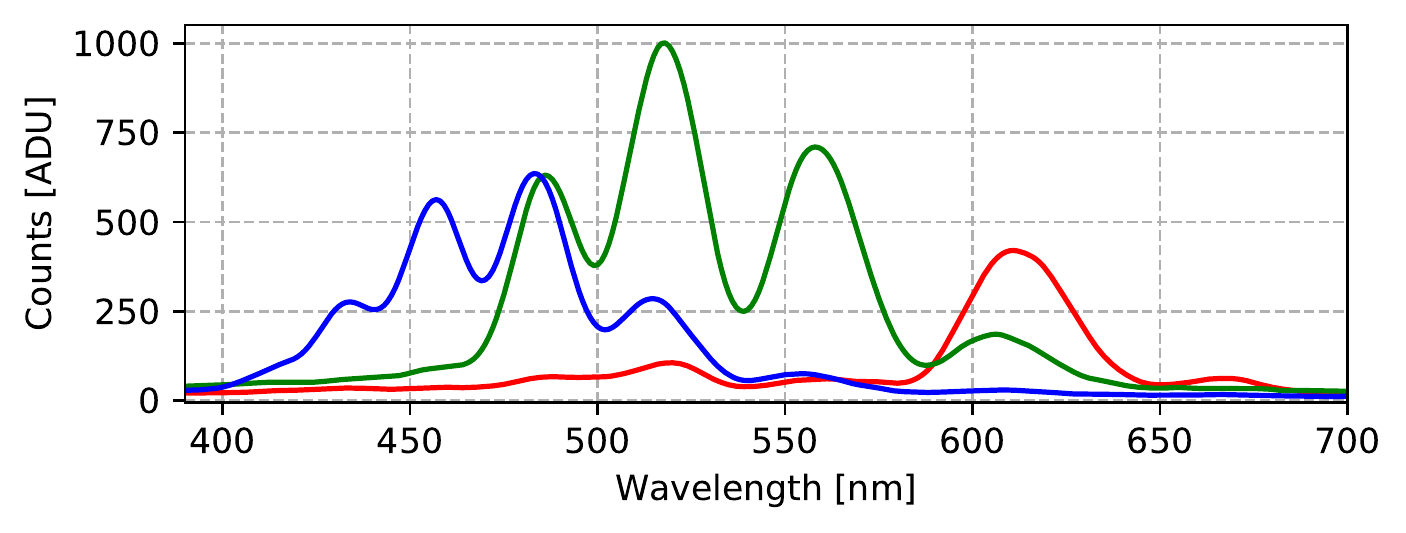}}
    \caption{Spectra from Fig.~\ref{f:raw_lines} after wavelength calibration.}
    \label{f:wavelength}
\end{figure}

The overall wavelength map is converted into individual wavelength maps for the RGBG$_2$ channels by de\-mosaicing it, as if it were an image itself. Finally, all rows are interpolated to the same 390--700 nm range in 1 nm steps, giving wavelength-calibrated spectra as shown in Fig.~\ref{f:wavelength}.

Since fluorescent lights are less ubiquitous than in the past, and will likely be fully replaced by LEDs in the foreseeable future, an alternative method may become necessary. For example, common features of SRFs such as the $\sim$580 nm edge in R bands \cite{Burggraaff2019SPECTACLE} may be used instead of spectral lines.

\subsection{Spectral Response Calibration} \label{ss:data:radiometry}

\begin{figure}[ht]
    \centering
    \subfloat[Fluorescent light. \label{f:SRF:TL}]{\includegraphics[width=8.2cm]{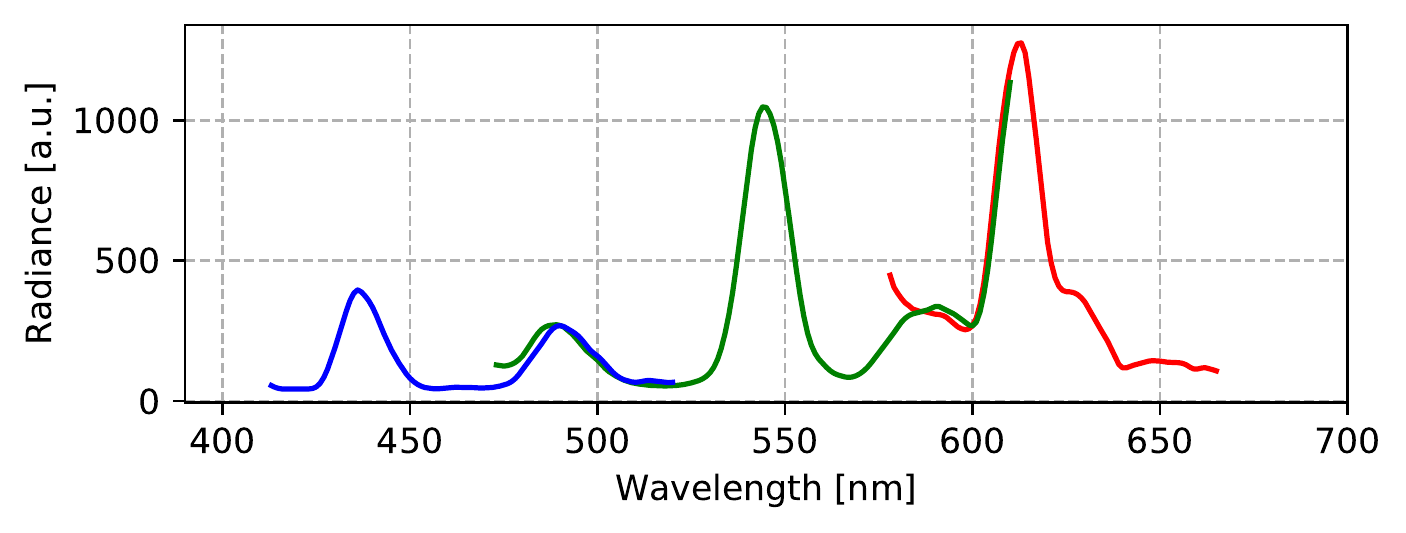}}
    \qquad
    \subfloat[Reflected sunlight through a 100\% polarizer. \label{f:SRF:100}]{\includegraphics[width=8.2cm]{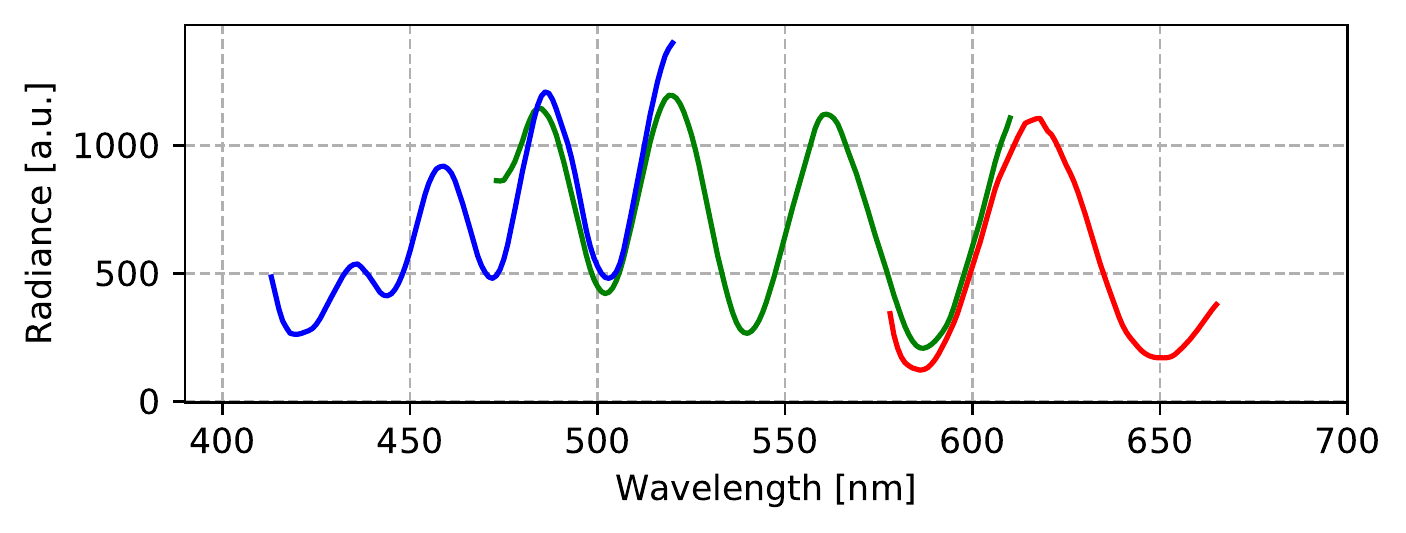}}
    \caption{Spectra from Fig.~\ref{f:wavelength} after spectral response calibration.}
    \label{f:SRF}
\end{figure}

The wavelength-calibrated spectra are corrected for the RGBG$_2$ spectral response functions (SRFs), again using SPECTACLE \cite{Burggraaff2019SPECTACLE}. The SRFs are interpolated to the same wavelengths as the data, after which the data are divided by the SRF. This gives radiances, a constant factor away from absolute radiometric units \cite{Burggraaff2019SPECTACLE}, as shown in Fig.~\ref{f:SRF}. To prevent amplifying noise and stray light, currently only wavelengths where the SRFs are $> 0.15$ (in relative units) are used. This restriction will be relaxed with better stray light reduction and correction. Even so, both spectra in Fig.~\ref{f:SRF} show excellent agreement between the RGB radiances. For fully unpolarized light measurements, this is the final calibration step.

For cameras without SPECTACLE SRFs, iSPEX 2 itself can be used to measure these using a known light source. This has been attempted with the original iSPEX using skylight reflected off white paper, with an RMS error of 5\% compared to reference data, increasing towards longer wavelengths \cite{Burggraaff2019SPECTACLE}. The skylight spectrum was model-based, introducing assumptions that cannot be tested in the same measurement. For this reason, it may be preferable to instead use consumer lamps, such as those characterized in the LICA database \cite{Tapia2018LICALamps}.

\subsection{Polarimetric Demodulation} \label{ss:data:polarimetry}

Finally, the DoLP and AoLP are retrieved by inverting Eq.~\eqref{e:spex} to demodulate the calibrated spectra. First, the retardance and polarimetric efficiency of the instrument must be calibrated using a known 100\% polarized light source \cite{Smit2019SPEXairborne}. For iSPEX 2, we plan to do this upon assembly using a rotating wiregrid polarizer. 

\begin{figure}[ht]
    \centering
    \subfloat[Fluorescent light. \label{f:retardance:TL}]{\includegraphics[width=8.2cm]{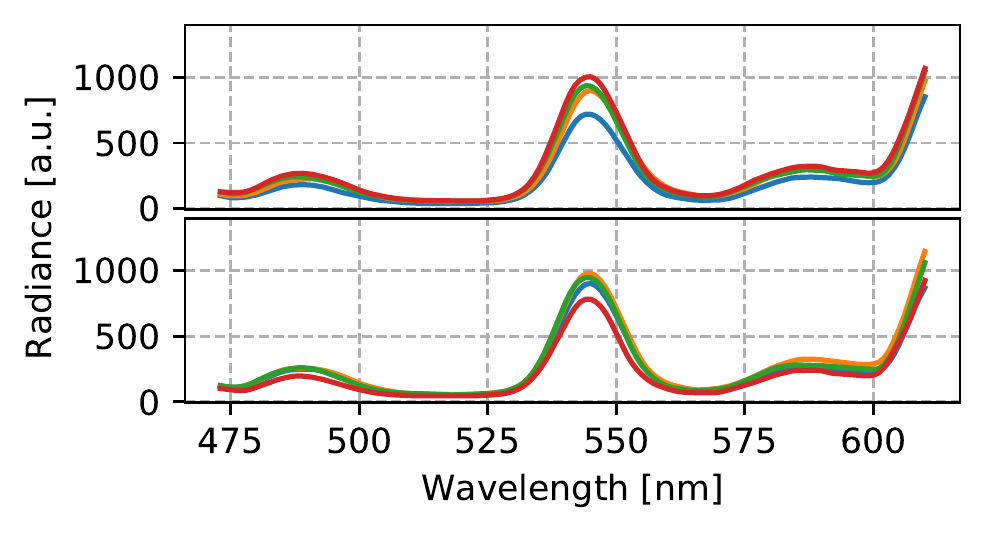}}
    \qquad
    \subfloat[Reflected sunlight through a 100\% polarizer. \label{f:retardance:100}]{\includegraphics[width=8.2cm]{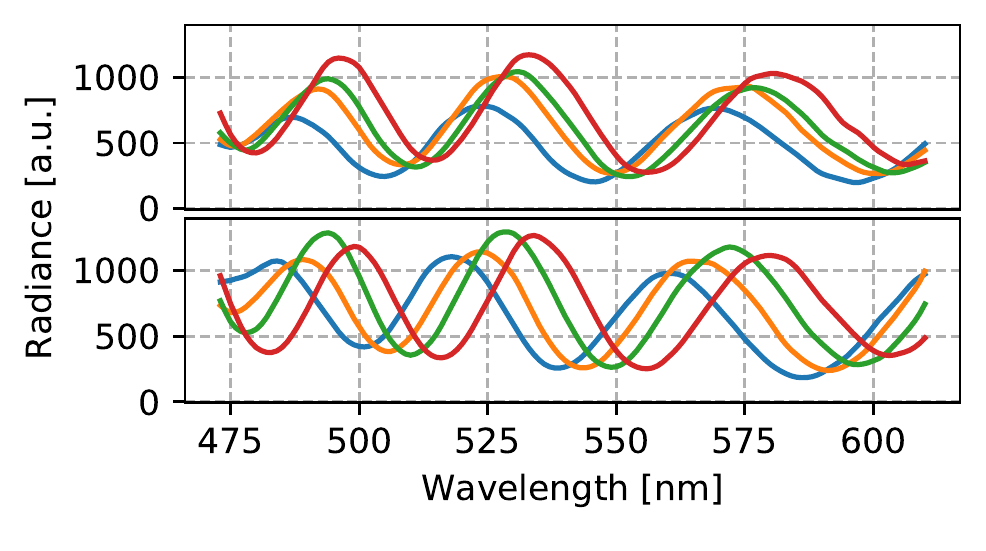}}
    \caption{Four rows in the G-band $+Q$ (top) and $-Q$ (bottom) spectra from Fig.~\ref{f:raw_data}. Each colored line corresponds to a row in the spectrum, though these are not the same between $+Q$ and $-Q$. Large variations are seen in the 100\% polarized spectrum but not the unpolarized fluorescent light spectrum.}
    \label{f:retardance}
\end{figure}

As can be seen in Fig.~\ref{f:retardance}, variations in retardance and efficiency exist along the $\pm Q$ slits of 3D-printed prototypes. These are partially due to fabrication issues, such as MOR foils flexing in their bracket due to wide production tolerances (Sec.~\ref{ss:production:print}), which will be resolved in the injection molded product. However, they are also partially due to issues including nonnormal incidence, since the optical path length and refractive index vary with the direction of propagation \cite{Hale1988retarders}; this will persist in the final product, necessitating a spatially dependent calibration of retardance and efficiency. We are currently characterizing these effects. A final complication is the fact that different sections of the slits see different targets, meaning inherent variations in DoLP and AoLP exist. The demodulation pipeline will have to account for this too.

The demodulation algorithm is in development. An iterative approach is likely necessary, fitting not only DoLP, AoLP, and the unpolarized spectrum $I_{in} (\lambda)$, but also instrumental effects including the QWP and MOR retardance, alignment and orientation of foils, the relative transmission between the two slits, and variations in the source spectrum along the slits \cite{Smit2019SPEXairborne, vanHarten2014groundSPEX}. The derived DoLP, AoLP, and $I_{in} (\lambda)$ are used to determine AOT for aerosol and $R_{rs}$ for ocean color measurements. The typical spectral resolution in DoLP and AoLP is approximately the modulation period \cite{Snik2009SPEXconcept} of 25--60 nm, though this can be lowered to the native spectral resolution of $\sim$9 nm (Sec.~\ref{ss:data:wavelength}), albeit with a lower accuracy \cite{vanHarten2014SPEXline}. The original iSPEX had a polarimetric accuracy (typical uncertainty) of $\approx 0.03$ in DoLP, mostly limited by defocus due to lacking camera controls \cite{Snik2014ispex}. Having solved these problems \cite{Burggraaff2019SPECTACLE}, we hope to increase the accuracy to sub-percent levels which enable retrieval of parameters such as effective radii and refractive indices (real and imaginary) \cite{DiNoia2015neuralnetworks}.

\section{FUTURE PERSPECTIVE} \label{s:future}

Production of iSPEX 2 is planned to start in the summer of 2020, and a number of calibration and validation efforts will take place around the same time. These are described here, along with current and future scientific applications.

\subsection{Calibration \& Validation} \label{s:future:validation}

Each iSPEX 2 unit will be factory-calibrated for retardance and polarimetric efficiency with a 100\% polarizer (Sec.~\ref{ss:data:polarimetry}) and a small number will be validated in the lab at various DoLP and AoLP using a glass plate setup \cite{Snik2014polreview}. This will allow for a thorough comparison in performance between iSPEX 2 and other sensors, as well as between iSPEX 2 units and between smartphones. The calibration data for each iSPEX 2 unit will be linked to its serial number in a database, from which the app will retrieve them.

iSPEX 2 AOT and $R_{rs}$ measurements will be validated through simultaneous observations with other instruments. For both, this will include groundSPEX, which is based on the same principle but with a much higher spectral resolution and a polarimetric accuracy of $\sim$1\% \cite{vanHarten2014groundSPEX}. AOT match-ups will also be done with MicroTOPS II, a handheld Sun photometer \cite{Morys2001MicroTOPS}, and AERONET \cite{Holben1998AERONET}. AERONET has previously been used to validate groundSPEX \cite{vanHarten2014groundSPEX}. $R_{rs}$ match-ups will be done using WISP-3 handheld and TriOS RAMSES shipborne spectroradiometers, similar to the HydroColor app \cite{Leeuw2018hydrocolor}. Both AOT and $R_{rs}$ validation measurements will largely take place within field campaigns organized through the MONOCLE consortium\footnote{\url{https://monocle-h2020.eu/}}.

\subsection{Scientific Applications} \label{s:future:science}

The main application of iSPEX 2 is as a low-cost instrument for citizen science measurements of air and water quality, using AOT and $R_{rs}$ as proxies. Both top-down and bottom-up approaches will be used for this. In the top-down approach, citizen scientists will be prompted by researchers to observe at a certain place or time, similar to the original iSPEX \cite{Snik2014ispex}. Conversely, in the bottom-up approach, citizen scientists can use iSPEX 2 independently, with researchers only providing support such as data processing and interpretation.

Planned top-down scientific applications of iSPEX 2 include high spatial resolution measurements of AOT and $R_{rs}$, extension of existing time series, and validation of satellite or airborne instruments. iSPEX 2 provides point measurements in arbitrary locations, facilitating extremely high spatial resolution. For example, a small group of citizen scientists standing along a lake shore can simultaneously map its reflectance (and thus its inherent properties) on meter scales. iSPEX 2 can also be used to fill in gaps in existing time series, for example if clouds prevented measurements during a satellite overpass. Finally, push notifications can be used to prompt citizen scientists to take validation measurements during a satellite overpass.

\subsection{Future Opportunities}

In addition to the currently planned applications, more experimental work with iSPEX 2 is also possible. For example, while we are currently focused on unpolarized $R_{rs}$, polarized $R_{rs}$ may provide additional information on water composition \cite{Gilerson2020spectropolarimetrywater, Neukermans2018remotesensingreview}. However, an optimal protocol for measuring polarized $R_{rs}$ will need to be found. Additionally, extending the principal plane measurements to the solar aureole, almucantar, and horizon may improve the AOT data and particle size distributions \cite{Vlemmix2010MAX-DOAS, Holben1998AERONET, Deepak1982aureole}. Outside remote sensing, iSPEX 2 will also be useful as a low-cost device for lab or field-going spectroscopy, for biological assaying and point-of-care diagnostics, among other purposes \cite{McGonigle2018spectrometers, Crocombe2018spectroscopy}.

Non-smartphone platforms also provide interesting opportunities. Unmanned aerial vehicles (UAVs) and webcams like the Raspberry Pi have cameras capable of professional-grade radiometry \cite{Burggraaff2019SPECTACLE}. Raspberry Pi-based systems could be used as low-cost autonomous field-going spectroradiometers. Already, UAVs with pushbroom spectrometers are delivering data products like $R_{rs}$ with high spatial and spectral resolution in a single fly-over \cite{Aasen2018UAVspectroscopy}. Using iSPEX 2, any camera can become a hyperspectral and polarimetric sensor.

\acknowledgments     
 
The authors wish to thank Akupara and Sanjana Panchagnula for providing pandemic-proof laboratory space.
Data analysis and visualization were done using the AstroPy, Matplotlib, NumPy, and SciPy libraries for Python.
This project has received funding from the European Union's Horizon 2020 research and innovation programme under grant agreement No 776480 (MONOCLE).


\bibliography{references} 
\bibliographystyle{spiebib} 

\end{document}